\documentclass[journal]{IEEEtran}

\usepackage{amsmath,amssymb,amsfonts,amsthm}
\newtheorem{theorem}{Theorem}
\newtheorem{proposition}[theorem]{Proposition}

\usepackage{algorithmic}
\usepackage{algorithm}
\usepackage{graphicx}
\graphicspath{{./figures/}}
\usepackage{textcomp}
\usepackage{xcolor}
\usepackage{cite}
\usepackage{multirow}
\usepackage{booktabs}
\usepackage{array}
\usepackage{bm}
\usepackage{url}
\usepackage{hyperref}
\usepackage{subcaption}
\usepackage{balance}
\makeatother
\makeatletter
\g@addto@macro\normalsize{%
  \setlength{\abovedisplayskip}{2pt}%
  \setlength{\belowdisplayskip}{2pt}%
  \setlength{\abovedisplayshortskip}{2pt}%
  \setlength{\belowdisplayshortskip}{2pt}%
}
\makeatother

\newcommand{\etal}{\textit{et al.}}

\begin{document}

\title{Identifiability-Guided Composite Load Modeling of AI Data Centers}

\author{
  Ding~Lin,
  ~Jianhui~Wang,~\IEEEmembership{Fellow,~IEEE}, Meng Yue,~\IEEEmembership{Member,~IEEE}, and Tianqiao Zhao,~\IEEEmembership{Member,~IEEE}
}

\markboth{IEEE Transactions on Power Systems}{Lin and Wang: Data Center Composite Load Modeling}

\maketitle

\begin{abstract}
The rapid growth of artificial-intelligence (AI) data centers is introducing large, workload-dependent loads that can affect the transient behavior of bulk power systems. Existing data-center load models incorporate increasingly detailed physical and workload dynamics, but provide limited guidance on how data-center-specific quantities should be
treated during calibration. This paper develops an identifiability-guided calibration framework for a branch-resolved model of IT, cooling, and auxiliary loads. The model exposes the workload-to-power law, cooling composition, UPS voltage gate, and reconnect-ramp duration as distinct calibration elements. A nuisance-projected Fisher-information criterion characterizes local identifiability of the cooling composition, while a workload non-equivalence result shows why voltage-only recalibration is insufficient for workload-dependent responses. These results guide the treatment of these elements within a grey-box estimator, with the learned workload law subsequently distilled into a closed form for transient-stability simulation. The framework is evaluated using 300,000 MIT Supercloud GPU power--utilization samples and 12,600 IEEE 39-bus fault scenarios under an event-grouped split. It achieves aggregate active-power $R^2=0.948$, compared with 0.856 for the best-performing learning baseline (LSTM). When trained on shallow voltage sags and tested on unseen deep sags, transient $R^2$ remains $0.750\pm0.014$, whereas the LSTM drops to $0.033\pm0.207$. Closed-loop simulations show close voltage-response agreement with a data-center-parameterized WECC PERC1 aggregate in 24 of 27 fault conditions, while the three protection-boundary cases lead to different retained-load states and voltage deviations of up to 0.33 p.u.
\end{abstract}


\vspace{-0.5cm}
\section{Introduction}
\label{sec:intro}

\IEEEPARstart{T}{he} rapid expansion of artificial intelligence workloads has made data centers one of the fastest-growing load categories on the North American bulk power system. Data-center electricity consumption in the United States reached approximately $4.4\%$ of total national demand in 2023 and is projected to grow to $6.7$--$12\%$ by 2028~\cite{shehabi20242024}. This growth is creating new reliability concerns for regional grids. The North American Electric Reliability Corporation  Large Loads Task Force has identified data centers as an emerging large-load category with implications for resource adequacy, operational balancing, and dynamic stability~\cite{force2025characteristics}. Recent field events further illustrate the potential system-level impact. On July~22, 2026, a transmission-line fault in Northern Virginia caused a cluster of data centers on the PJM interconnection to transfer to backup power within roughly thirty seconds, removing approximately $3$~GW of demand. The resulting voltage and frequency disturbance propagated across the Eastern Interconnection and required nearly ten minutes to stabilize~\cite{pjm2026disconnect}.

From the perspective of dynamic load modeling, data centers differ from conventional aggregate loads in several important respects. GPU-intensive AI workloads can produce sub-second power variations directly shaped by software-level scheduling and may interact with local and inter-area oscillatory modes~\cite{ko2025wide}. Uninterruptible power-supply (UPS) protection introduces discontinuous behavior. When the point of
common coupling (PCC) voltage falls below a protection threshold, IT load may transfer away from the grid and then reconnect through a staged ramp~\cite{jimenez2025data}. Cooling systems further combine motor-driven and power-electronic loads with dynamics spanning electrical and thermal timescales~\cite{li2025electronics}. These mechanisms are not well captured by a single aggregate representation and motivate load models that preserve the physically distinct IT, cooling, and auxiliary responses relevant to transient studies.

Recent studies and industry efforts have begun to address these issues through more detailed representations of data-center dynamics. Jimenez-Ruiz and Milano~\cite{jimenez2025data} developed a dynamic data-center model incorporating UPS behavior, induction-motor cooling, and workload-driven power variations for power-system stability analysis. Gyang \etal{}~\cite{gyang2025} presented an integrated electrical and thermal dynamic model of a data center for stability studies. Li and Li~\cite{li2025electronics} examined AI-load dynamics from a power-electronics perspective and characterized the slew-rate constraints imposed by multi-stage power-conversion architectures. Zhao and Zhao~\cite{zhao2026dynamic} developed a component-informed model of the data-center power-delivery system to investigate interactions between converter dynamics and power-system oscillatory modes. On the standards side, the WECC Modeling and Validation Subcommittee has recommended the PERC1 (Power Electronic Reconnecting and Ceasing) model as a minimum-fidelity aggregate representation for computational loads, capturing voltage-dependent load ceasing and reconnection behavior~\cite{weber2026perc1}. As these representations incorporate more workload, protection, and component-level dynamics, reliable calibration of the resulting models becomes increasingly important.

Data-driven and hybrid load models offer a complementary route to calibration. Neural networks can reproduce nonlinear dynamic responses with high accuracy~\cite{arif2017load}, but purely black-box representations do not directly expose physically interpretable model quantities or provide a natural path to conventional transient-stability models. Symbolic regression can produce closed-form expressions~\cite{lin2026reinforcement}, while physics-informed learning can preserve selected physical constraints~\cite{huang2022applications}. Kolmogorov-Arnold Networks (KANs) provide another means of learning low-dimensional nonlinear relations through trainable univariate functions~\cite{liu2025kan}. Lu \etal{}~\cite{lu2026} recently combined a parameterized
physics-based structure with data-driven calibration for
data-center load identification.
These methods support flexible calibration within different
model representations, but their use still requires determining
which physical quantities can be reliably recovered from the
available measurements.

This issue is closely related to parameter identifiability in composite load modeling. Previous studies have used trajectory sensitivities, Fisher information, and related methods to assess which parameters of conventional composite load models can be inferred from terminal measurements~\cite{arif2017load}. Such analyses have largely focused on conventional components, including induction motors, ZIP loads, and single-phase air-conditioning. Data centers introduce workload-, protection-, and cooling-related
quantities with different degrees of recoverability from available
electrical and facility measurements. Treating all of these
quantities as freely estimable may therefore produce weakly
identified or physically confounded parameterizations.
The resulting calibration question is which quantities can be
reliably estimated from the available measurements, and which
should be retained in prescribed form or externally anchored.

Motivated by this gap, this paper develops an identifiability-guided calibration framework for data-center dynamic load modeling. Rather than treating all data-center-specific quantities as equally estimable, the framework differentiates their calibration roles according to recoverability under the available PCC measurements and facility telemetry. The main contributions are:

\begin{enumerate}

\item[1)] We formulate a branch-resolved composite load representation for
AI data centers that separates IT, cooling, and auxiliary dynamics while
exposing the workload-to-power relation, cooling composition, UPS voltage
gate, and reconnect-ramp duration as distinct calibration elements. This
representation enables these elements to be treated separately during
calibration rather than absorbed into a single aggregate response.

\item[2)] We establish a recoverability-based approach for selecting the
calibration treatment of each exposed calibration element. A nuisance-projected
Fisher-information criterion characterizes local identifiability of the
cooling composition in the presence of nuisance parameters, while a
workload non-equivalence result shows why voltage-only recalibration is
insufficient for workload-dependent responses. Together with empirical
recoverability tests, these results provide the basis for determining which
elements are estimated, retained in prescribed form, or externally anchored.

\item[3)] We develop an identifiability-guided grey-box estimator that combines
branch-level supervision with a two-phase training strategy. Branch-level
supervision separates terminal fitting from sub-component recovery, while
two-phase training separates joint model training from subsequent refinement
of the learnable workload law and cooling share. Post-training symbolic distillation
converts the learned workload relation into a closed form suitable for
transient-stability simulation.

\end{enumerate}

Tests on $12{,}600$ IEEE~39-bus fault scenarios show that terminal accuracy
does not guarantee recovery of the underlying cooling dynamics: aggregate-only training preserves
the PCC active-power response but fails to recover the cooling-branch response
and composition. When trained on shallow voltage sags and tested on unseen
deep sags, the proposed model retains a transient $R^2$ of $0.750$, compared
with $0.033$ for the LSTM. In closed-loop simulations, the proposed model
agrees closely with the PERC1 aggregate in $24$ of
$27$ fault conditions, while the three protection-boundary cases lead to different retained-load states. Detailed derivations, model parameter tables, and evaluation
protocols are available at
\url{https://github.com/dinglin613/DCLR}.

\vspace{-0.5cm}
\section{Data Center Composite Load Model}
\label{sec:model}

This section presents the physics-structured composite load model. The facility power is decomposed into physically distinct IT, cooling,
and auxiliary branches, with the associated calibration elements defined below.
\vspace{-0.5cm}
\subsection{Facility Decomposition}
The model is formulated for phasor-domain electromechanical
transient-stability studies and represents the data-center facility
as a composite load at the PCC.

The facility active and reactive powers at the PCC are
\begin{align}
P_{\mathrm{DC}}(t) &= P_{\mathrm{IT}}(t) + P_{\mathrm{cool}}(t) + P_{\mathrm{aux}}(t), \label{eq:p_decomp}\\
Q_{\mathrm{DC}}(t) &= Q_{\mathrm{IT}}(t) + Q_{\mathrm{cool}}(t) + Q_{\mathrm{aux}}(t), \label{eq:q_decomp}
\end{align}
where the subscripts denote IT, cooling, and auxiliary loads.
\vspace{-0.2cm}
\subsection{IT Load and UPS Protection}
\label{sec:model_it}

The IT load is a constant-power load (CPL) served through a UPS. Its PCC-side active power is
\begin{equation}
P_{\mathrm{IT}}(t) = \gamma(V,t)\, P_{\mathrm{rated}}\, \alpha_{\mathrm{IT}}\, g(w(t)),
\label{eq:p_it}
\end{equation}
where $V$ is the PCC voltage magnitude in p.u., $P_{\mathrm{rated}}$ is the facility active-power base, and {\color{black}$\alpha_{\mathrm{IT}}$ is the nominal IT share} at the reference operating condition $(w_{\mathrm{ref}}=1,\,V=1~\mathrm{p.u.})$. $g:[0,1]\rightarrow\mathbb{R}_+$ is the workload-to-power function normalized so that $g(w_{\mathrm{ref}})=1$, $w(t)$ is the GPU utilization aggregate, and $\gamma(V,t)\in[0,1]$ is the UPS connection factor. The UPS rectifier is represented by a fixed near-unity power factor
$\mathrm{pf}_{\mathrm{IT}}$, giving
$Q_{\mathrm{IT}}
=
P_{\mathrm{IT}}
\tan\!\bigl(\cos^{-1}\mathrm{pf}_{\mathrm{IT}}\bigr)$.
For a fixed workload, the connected IT branch is modeled as constant power
away from the protection-transition region, such that
$\partial P_{\mathrm{IT}}/\partial V\approx0$.

The UPS connection factor combines a voltage gate and a reconnect factor:
\begin{equation}
\gamma(V,t) = \gamma_h(V)\cdot r(t),\qquad \gamma_h(V)\triangleq\sigma\!\left(\frac{V-V_{\mathrm{th}}}{\epsilon}\right),
\label{eq:ups_gamma}
\end{equation}
where $\sigma(\cdot)$ is the sigmoid, {\color{black}$V_{\mathrm{th}}$ is the UPS transfer threshold}, and $\epsilon$ is the gate-smoothing width. Here $r(t)$ denotes the general reconnect factor. In the nominal physical
model, $r(t)=r_{\mathrm{nom}}(t;\tau_{\mathrm{ramp}})$. The disconnection and reconnect-eligibility times are
\begin{align}
t_{\mathrm{disc}} &\triangleq \inf\{t:\, V(t)<V_{\mathrm{th}}\},\\
t_{\mathrm{re}} &\triangleq \inf\{t>t_{\mathrm{disc}}:\, V(t)\ge V_{\mathrm{th}}+\Delta V\},
\label{eq:t_re}
\end{align}
where $\Delta V$ is the reconnect hysteresis margin. The nominal reconnect factor is
\begin{equation}
r_{\mathrm{nom}}(t;\tau_{\mathrm{ramp}})=\begin{cases}
1, & t<t_{\mathrm{disc}}\\
0, & t_{\mathrm{disc}}\le t<t_{\mathrm{re}}\\
\min\!\bigl(1,(t-t_{\mathrm{re}})/\tau_{\mathrm{ramp}}\bigr), & t\ge t_{\mathrm{re}}
\end{cases}
\label{eq:ramp}
\end{equation}
If disconnection is not triggered, $r_{\mathrm{nom}}(t;\tau_{\mathrm{ramp}})\equiv 1$; if disconnection occurs but the recovery condition is not met, $r_{\mathrm{nom}}(t;\tau_{\mathrm{ramp}})$ remains zero for $t\ge t_{\mathrm{disc}}$. 

\vspace{-0.3cm}
\subsection{Cooling Load and Motor Composition}
\label{sec:model_cool}

The cooling branch represents the electrical demand of chiller compressors, cooling towers, and computer-room air-conditioning units. The associated motor loads are represented as a mixture of direct-on-line
(DOL) induction motors and motors supplied through variable-frequency
drives (VFDs):
\begin{align}
P_{\mathrm{cool}}(t) &= \chi_{\mathrm{cool}}(t)\,P_{\mathrm{rated}}\,\alpha_{\mathrm{cool}}\,\Pi_P(V,s;\rho)\, f_{\mathrm{th}}(T_{\mathrm{room}}), \label{eq:p_cool}\\
\Pi_P(V,s;\rho) &\triangleq \rho\,\bar P_{\mathrm{IM}}(V,s) + (1-\rho)\,P_{\mathrm{VFD}}(V).
\end{align}
Here {\color{black}$\alpha_{\mathrm{cool}}$ is the nominal cooling share} on the same facility power base, and $\rho\in[0,1]$ is the DOL fraction (with $1-\rho$ the VFD fraction). The functions $\bar P_{\mathrm{IM}}(V,s)$ and $P_{\mathrm{VFD}}(V)$ are the normalized active-power characteristics of the DOL and VFD components, respectively. Cooling demand is modulated by the thermal factor
\begin{equation}
f_{\mathrm{th}}(T_{\mathrm{room}}) = 1 + k_{\mathrm{th}}\,\frac{T_{\mathrm{room}}-T_{\mathrm{room},0}}{\Delta T_{\mathrm{ref}}},
\label{eq:fth}
\end{equation}
where $k_{\mathrm{th}}$ is the thermal-factor gain, $T_{\mathrm{room},0}$ is the pre-disturbance room-temperature setpoint, and $\Delta T_{\mathrm{ref}}$ is the temperature normalization scale. The relay state $\chi_{\mathrm{cool}}(t)\in\{0,1\}$ applies to both components and latches from $1$ to $0$ once the terminal voltage remains below the relay threshold for the specified dwell time and the relay delay has elapsed. The VFD component additionally drops out below its ride-through voltage $V_{\mathrm{d,VFD}}$ and re-engages when $V\ge V_{\mathrm{d,VFD}}$.

The induction motor is represented as a single-cage equivalent circuit with stator and rotor resistances $R_s, R_r$, leakage reactances $X_s, X_r$, and magnetizing reactance $X_m$. The slip is $s=(\omega_s-\omega_r)/\omega_s$, where $\omega_s$ and $\omega_r$ are the synchronous and rotor speeds; the motor equations use $\omega_s\simeq 1$~p.u.\ over the simulated transient interval. The magnetizing reactance is included as a shunt term in the reactive-power expression, consistent with the WECC composite load representation~\cite{wecc_clm}.

The squared series-branch current magnitude and series-branch active power are
\begin{align}
I_{\mathrm{sr}}^2 &= \frac{V^2}{(R_s+R_r/s)^2+(X_s+X_r)^2}, \label{eq:is_im}\\
P_{\mathrm{IM}}^{\mathrm{raw}} &= I_{\mathrm{sr}}^2\,(R_s + R_r/s). \label{eq:pim}
\end{align}
The normalized motor active-power characteristic is
$\bar P_{\mathrm{IM}}(V,s)
=\kappa_{\mathrm{IM}}P_{\mathrm{IM}}^{\mathrm{raw}}(V,s)$, where
$\kappa_{\mathrm{IM}}
=1/P_{\mathrm{IM}}^{\mathrm{raw}}(1,s_{\mathrm{op}})$ and
$s_{\mathrm{op}}$ is the nominal operating slip at $V=1$~p.u.
This normalization gives
$\bar P_{\mathrm{IM}}(1,s_{\mathrm{op}})=1$.

Slip evolves according to the rotor swing equation
\begin{equation}
2H\frac{ds}{dt} = T_{\mathrm{mech}}(T_{\mathrm{room}}) - T_{\mathrm{em}}(V,s),
\label{eq:slip}
\end{equation}
where $H$ is the aggregate cooling-motor inertia. {\color{black}The torques are $T_{\mathrm{em}}(V,s)=I_{\mathrm{sr}}^2R_r/s$ and $T_{\mathrm{mech}}(T_{\mathrm{room}})=T_{\mathrm{m},0}f_{\mathrm{th}}(T_{\mathrm{room}})$.} The pre-fault initial slip $s_0$ is the low-slip root of the torque balance
\begin{equation}
T_{\mathrm{em}}(V_0,s_0) = T_{\mathrm{mech}}(T_{\mathrm{room},0}),
\label{eq:slip_equilibrium}
\end{equation}
where $V_0$ is the pre-fault terminal voltage; when $V_0=1$~p.u., the equilibrium solution satisfies $s_0=s_{\mathrm{op}}$. 

The room temperature follows the normalized energy balance in deviation form
\begin{equation}
\tau_{\mathrm{th}}\frac{d\tilde T_{\mathrm{room}}}{dt} = \bigl(\tilde P_{\mathrm{IT}}(t) - \tilde P_{\mathrm{IT}}^{\,\mathrm{nom}}\bigr) - \mathrm{COP}\cdot\bigl(\tilde P_{\mathrm{cool}}(t) - \tilde P_{\mathrm{cool}}^{\,\mathrm{nom}}\bigr).
\label{eq:thermal}
\end{equation}
Here $\tilde T_{\mathrm{room}}=(T_{\mathrm{room}}-T_{\mathrm{room},0})/\Delta T_{\mathrm{ref}}$ is the normalized temperature deviation, and $\tilde P_k = P_k/P_{\mathrm{rated}}$ is the per-unit branch power on the facility base for $k\in\{\mathrm{IT},\mathrm{cool}\}$, with pre-fault values $\tilde P_k^{\,\mathrm{nom}} = P_k(t_0^-)/P_{\mathrm{rated}}$ evaluated at the instant $t_0^-$ immediately before the disturbance. $\mathrm{COP}$ denotes the cooling coefficient of performance, and {\color{black}$\tau_{\mathrm{th}}$ is the aggregate room thermal time constant.}

In the physical simulation, $T_{\mathrm{room}}$ evolves according to Eq.~\eqref{eq:thermal}; during identification, the measured facility temperature $\hat T_{\mathrm{room}}$ is supplied to the estimator. The cooling-branch reactive power uses the same DOL/VFD composition:
\begin{align}
Q_{\mathrm{cool}}(t) &= \chi_{\mathrm{cool}}(t)\,P_{\mathrm{rated}}\,\alpha_{\mathrm{cool}}\,\Pi_Q(V,s;\rho)\, f_{\mathrm{th}}(T_{\mathrm{room}}), \label{eq:q_cool}\\
\Pi_Q(V,s;\rho) &\triangleq \rho\,\bar Q_{\mathrm{IM}}(V,s) + (1-\rho)\,Q_{\mathrm{VFD}}(V),
\end{align}
where $\bar Q_{\mathrm{IM}}(V,s) = \kappa_{\mathrm{IM}}\bigl[I_{\mathrm{sr}}^2(X_s+X_r) + V^2/X_m\bigr]$ is the normalized induction-motor reactive-power characteristic and $Q_{\mathrm{VFD}}(V)$ is the VFD displacement-reactive characteristic. {\color{black}On the cooling-branch base, $P_{\mathrm{VFD}}(V)=\mathbf{1}\{V\ge V_{\mathrm{d,VFD}}\}$ and $Q_{\mathrm{VFD}}(V)=q_{\mathrm{disp}}P_{\mathrm{VFD}}(V)$.}

\vspace{-0.2cm}
\subsection{Auxiliary Load}

Auxiliary loads, including lighting, fire suppression, and security systems, are represented by a ZIP polynomial:
\begin{align}
P_{\mathrm{aux}}(t) &= P_{\mathrm{rated}}\,\alpha_{\mathrm{aux}}\,\bigl(k_Z\,V(t)^2 + k_I\,V(t) + k_P\bigr), \label{eq:p_aux}\\
Q_{\mathrm{aux}}(t) &= P_{\mathrm{aux}}(t)\,\tan\!\big(\cos^{-1}\mathrm{pf}_{\mathrm{aux}}\big), \label{eq:q_aux}
\end{align}
where {\color{black}$\alpha_{\mathrm{aux}}$ is the nominal auxiliary share on the facility power base.} The ZIP coefficients satisfy $k_Z+k_I+k_P=1$, and the nominal branch shares satisfy $\alpha_{\mathrm{IT}}+\alpha_{\mathrm{cool}}+\alpha_{\mathrm{aux}}=1$.
\vspace{-0.2cm}
\subsection{Identification Problem}
\label{sec:model_identification}

Let
\begin{equation}
\bm{x}_t = [V_t,f_t,w_t,\hat T_{\mathrm{room},t}]^\top,\quad
\bm{y}_t = [P_{\mathrm{DC},t},Q_{\mathrm{DC},t}]^\top,
\end{equation}
where $(V_t,f_t)$ are measured at the PCC and $(w_t,\hat T_{\mathrm{room},t})$ are facility telemetry. Given a length-$L$ input window and the physical state at its initial sample, the estimation problem is to recover $\bm{y}_t$ via
\begin{equation}
\hat{\bm{y}}_t = \mathcal{F}\!\left(\bm{x}_{t-L+1:t},\,\bm{\xi}_{t-L+1};\,\bm{\Theta}\right),
\label{eq:problem}
\end{equation}
where $\bm{\xi}$ collects the internally propagated physical states,
including the motor slip $s$, the cooling-relay
$\chi_{\mathrm{cool}}$ and its dwell/delay timers, and the UPS connection
and reconnect-ramp states. The set $\bm{\Theta}$ collects the
estimator parameters.

\section{Identifiability-Guided Estimator Framework}
\label{sec:framework}
This section develops the estimator for the model of Section~\ref{sec:model}. The calibration treatment and its analytical basis are first established, followed by the estimator architecture, training, and deployment summarized in Fig.~\ref{fig:architecture}.
{
\setlength{\abovecaptionskip}{0pt}  
\setlength{\belowcaptionskip}{-15pt} 
\begin{figure*}[t]
\centering
\includegraphics[width=0.75\linewidth, height=0.41\linewidth]{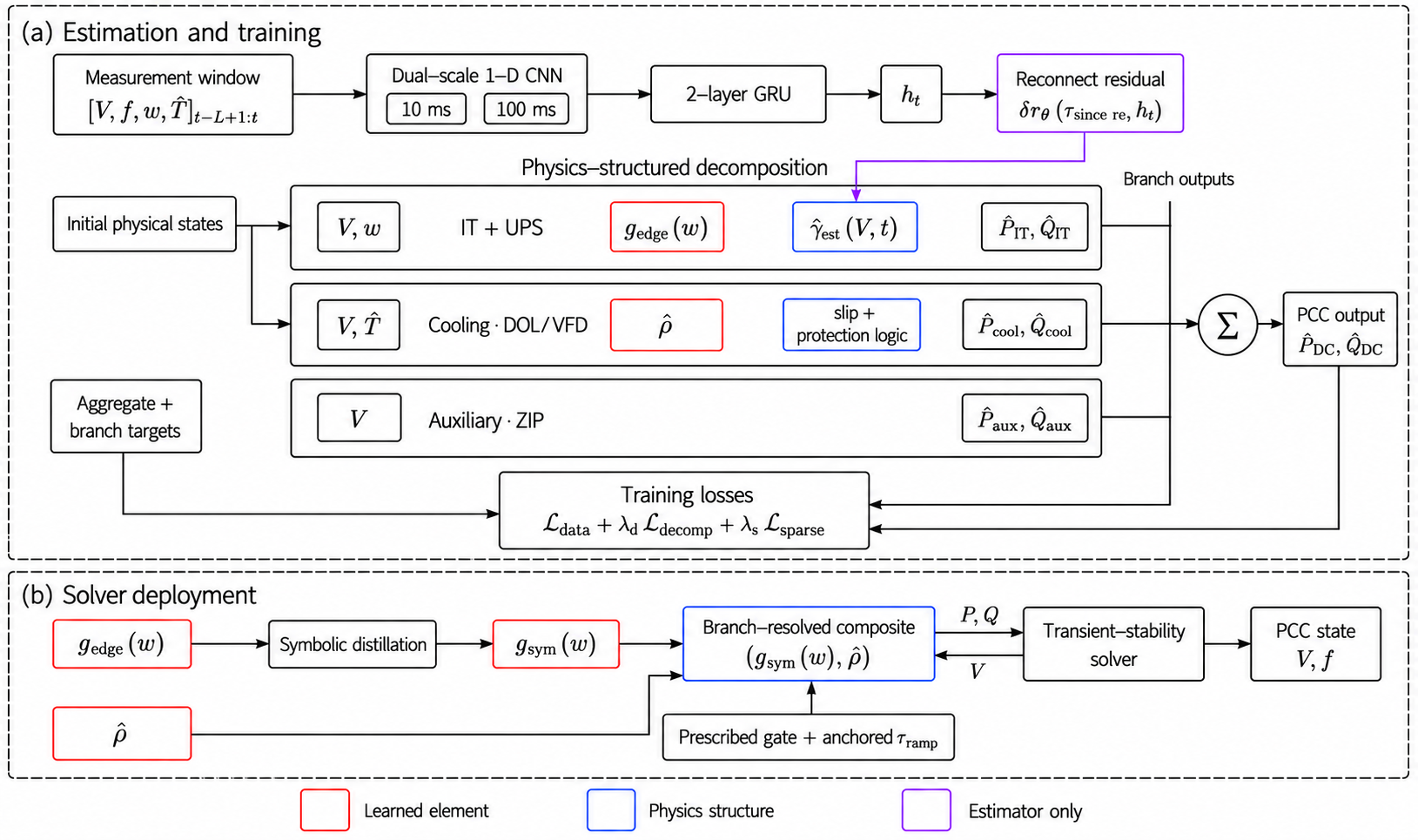}
\caption{Proposed framework: (a) grey-box estimator and training objective; (b) symbolic distillation and solver deployment.}
\label{fig:architecture}
\end{figure*}}

\vspace{-0.4cm}
\subsection{Available Information and Calibration Treatment}
\label{sec:framework_placement}
Here, calibration elements are exposed functional relations or
low-dimensional parameters treated separately during calibration and
deployment: $g(w)$, $\rho$, $\gamma_h(V)$, and $\tau_{\mathrm{ramp}}$.
The internally propagated states $\bm\xi$ are not calibration elements.
Branch powers $(P_k,Q_k)$ serve as additional training targets but are
not required at inference. These targets are supplied by the simulator
in this study. Field calibration with the same objective would require
branch-level $P/Q$ measurements or validated branch estimates.

Table~\ref{tab:identif} summarizes the treatments and their basis.
Theorem~\ref{thm:rho} characterizes the local identifiability of $\rho$,
while Proposition~\ref{prop:cmpldw} establishes the need for workload
information. The prescribed protection structure and the empirical
tests in Section~\ref{sec:case_interface} support the remaining treatments.

\begin{table}[t]
\centering
\caption{Calibration elements, assigned treatments, and supporting basis
under the available training information.}
\label{tab:identif}
\footnotesize
\setlength{\tabcolsep}{3pt}
\renewcommand{\arraystretch}{1.15}
\begin{tabular}{@{}p{1.7cm}p{2.5cm}p{4.5cm}@{}}
\toprule
Element & Treatment & Basis \\
\midrule

$g(w)$ workload law
&
Learn and symbolically distill
&
Observed workload variation and training-time branch supervision;
workload-input necessity from Proposition~\ref{prop:cmpldw}
\\

$\rho$ DOL/VFD share
&
Estimate as a scalar
&
Nuisance-projected criterion with an anchored cooling scale
(Theorem~\ref{thm:rho}); numerical Fisher-information analysis and controlled recovery
\\

$\gamma_h(V)$ UPS voltage gate
&
Retain in estimator and deployment
&
Prescribed voltage-dependent protection law with externally specified
threshold and smoothing
\\

$\tau_{\mathrm{ramp}}$ reconnect-ramp duration
&
Anchor for deployment; estimator-only residual $\delta r_\theta$
&
The residual-derived equivalent timescale does not track the imposed ramp
time within the present estimator
\\

\bottomrule
\end{tabular}
\end{table}

\vspace{-0.3cm}
\subsection{Analytical Basis for Calibration Treatment}
\label{sec:framework_id}

\emph{1) Cooling-composition identifiability.}

\begin{theorem}[Nuisance-projected local identifiability of the DOL/VFD share]
\label{thm:rho}
Let the stacked cooling-branch observations follow
\begin{equation}
\bm{y} \sim \mathcal{N}\bigl(\bm{\mu}(\rho,\bm{\eta}),\,R\bigr),\qquad R\succ 0,
\label{eq:cool_obs}
\end{equation}
where $\rho$ is the DOL/VFD share, $\bm{\eta}$ collects the motor, VFD, and initial-state nuisance parameters, and $R$ is known and independent of $(\rho,\bm{\eta})$. The prescribed PCC-voltage trajectory $V$ and the facility temperature telemetry $\hat T_{\mathrm{room}}$ are observed, the cooling branch remains connected throughout the observation window ($\chi_{\mathrm{cool}}\equiv 1$), the time-varying cooling-branch scale $C_k = P_{\mathrm{rated}}\,\alpha_{\mathrm{cool}}\,f_{\mathrm{th}}(\hat T_{\mathrm{room},k})$ is known or externally anchored, and $\bm{\mu}$ is continuously differentiable in a neighborhood of $(\rho_0,\bm{\eta}_0)$. Define the whitened sensitivity blocks
\begin{equation}
\bm{j}_\rho = R^{-1/2}\,\frac{\partial\bm{\mu}}{\partial\rho}\bigg|_{(\rho_0,\bm{\eta}_0)},\qquad
J_\eta = R^{-1/2}\,\frac{\partial\bm{\mu}}{\partial\bm{\eta}}\bigg|_{(\rho_0,\bm{\eta}_0)}.
\label{eq:whitened}
\end{equation}
The Fisher information for $\rho$ remaining after projection onto the nuisance-parameter directions is
\begin{align}
\mathcal{I}_{\rho\mid\eta}
&= F_{\rho\rho} - F_{\rho\eta}F_{\eta\eta}^{\dagger}F_{\eta\rho} \notag\\
&= \bigl\|(I-J_\eta J_\eta^{\dagger})\,\bm{j}_\rho\bigr\|_2^{\,2},
\label{eq:info_rho}
\end{align}
where $F_{\rho\rho}=\bm{j}_\rho^\top\bm{j}_\rho$, $F_{\rho\eta}=\bm{j}_\rho^\top J_\eta$, $F_{\eta\rho}=F_{\rho\eta}^\top$, $F_{\eta\eta}=J_\eta^\top J_\eta$, $I$ is the identity matrix, and $(\cdot)^{\dagger}$ denotes the Moore--Penrose pseudoinverse. The second form remains meaningful when $F_{\eta\eta}$ is rank deficient. Then $\rho$ is locally identifiable to first order in the presence of the
nuisance parameters if and only if
\begin{equation}
\mathcal{I}_{\rho\mid\eta}>0,\qquad\text{equivalently}\qquad \bm{j}_\rho\notin\mathrm{col}(J_\eta),
\end{equation}
and any regular locally unbiased estimator satisfies $\mathrm{var}(\hat\rho)\ge\mathcal{I}_{\rho\mid\eta}^{-1}$.
\end{theorem}
{\color{black}\noindent\emph{Proof.} At $(\rho_0,\bm\eta_0)$, let $\bm z=R^{-1/2}(\bm y-\bm\mu)$ and $P_\eta=J_\eta J_\eta^\dagger$. The efficient score for $\rho$ is $S_{\mathrm{eff}}=[(I-P_\eta)\bm j_\rho]^\top\bm z$, with variance $\|(I-P_\eta)\bm j_\rho\|_2^2$. Since $P_\eta=J_\eta(J_\eta^\top J_\eta)^\dagger J_\eta^\top$, expansion gives~\eqref{eq:info_rho}; this information is positive exactly when $\bm j_\rho\notin\mathrm{col}(J_\eta)$. A nonzero first-order change in $\rho$ can be canceled by nuisance perturbations exactly when $\bm j_\rho\in\mathrm{col}(J_\eta)$, establishing the identifiability criterion. Local unbiasedness with respect to $(\rho,\bm\eta)$ gives $\operatorname{cov}(\hat\rho,\bm j_\rho^\top\bm z)=1$ and $\operatorname{cov}(\hat\rho,J_\eta^\top\bm z)=\bm0$, hence $\operatorname{cov}(\hat\rho,S_{\mathrm{eff}})=1$. Cauchy--Schwarz then yields $\operatorname{var}(\hat\rho)\mathcal I_{\rho\mid\eta}\ge1$, giving the variance bound when $\mathcal I_{\rho\mid\eta}>0$. \hfill$\square$\par}
Branch-resolved supervision supplies the cooling observations assumed
above. Under aggregate-only identification, the IT and auxiliary
contributions would enter as additional nuisance quantities.

When the nuisance parameters are held fixed, the information for
$\rho$ reduces to $F_{\rho\rho}$. For $M$ samples with measurement
errors independent across samples and between the $P$ and $Q$
channels, this becomes
\begin{equation}
\mathcal{I}_{\rho}=F_{\rho\rho}
=\sum_{k=1}^{M} C_k^2
\left(
\frac{\Delta P_k^2}{\sigma_P^2}
+\frac{\Delta Q_k^2}{\sigma_Q^2}
\right),
\label{eq:info_rho_diag}
\end{equation}
where $\sigma_P^2$ and $\sigma_Q^2$ are the $P$- and $Q$-channel
measurement-noise variances,
$\Delta P_k=\bar P_{\mathrm{IM},k}-P_{\mathrm{VFD},k}$, and
$\Delta Q_k=\bar Q_{\mathrm{IM},k}-Q_{\mathrm{VFD},k}$.
This special case illustrates how differences between the DOL and VFD
component responses over the observation window provide information
about $\rho$. With unknown nuisance parameters, recoverability is
instead governed by the nuisance-projected sensitivity in
  Eq.~\eqref{eq:info_rho}.

\emph{2) Workload-input necessity.}

\begin{proposition}[Workload-induced non-equivalence under matched grid-side inputs]
\label{prop:cmpldw}
Let the workload-modulated IT trajectory be
$P_{\mathrm{IT}}(V,t,w) = S_{\mathrm{IT}}\,\Phi[V](t)\,g(w(t))$,
where $S_{\mathrm{IT}}\triangleq P_{\mathrm{rated}}\,\alpha_{\mathrm{IT}}$ is the nominal IT scale and $\Phi[V](t)\in[0,1]$ is any prescribed UPS connection factor determined by the voltage history under fixed protection settings and initial connection state. Let $\mathcal{P}$ denote a model family parameterized by $\theta_P$,
whose output $P^{\mathcal{P}}(\bm z(\cdot);\theta_P)$ depends only on
the grid-side input history $\bm z(t)=[V(t),f(t)]$. For any two scenarios
$\{(\bm z(t),w_1(t)),(\bm z(t),w_2(t))\}$ that share the same grid-side
input trajectory but differ in workload, and for any single fixed
parameter vector $\theta_P$ applied to both,
\begin{multline}
\sum_{i=1}^{2}\int_0^T\!\!\big[P^{\mathcal{P}}(\bm{z}(\cdot);\theta_P)(t) - P_{\mathrm{IT}}(V,t,w_i)\big]^2 dt \\
\;\geq\; \frac{S_{\mathrm{IT}}^2}{2}\int_0^T\!\!\Phi[V](t)^2\,\big[g(w_1(t)) - g(w_2(t))\big]^2\,dt.
\label{eq:perc1_lower}
\end{multline}
Consequently, whenever there exists a subinterval of $[0,T]$ on which
$\Phi[V](t)>0$ and $g(w_1(t))\neq g(w_2(t))$, no single fixed
parameterization of $\mathcal{P}$ can fit both scenarios simultaneously,
and the joint error is bounded away from zero.
\end{proposition}
{\color{black}\noindent\emph{Proof.} With the same grid-side history, a fixed grid-only model produces the same prediction $\bar P(t)$ in both scenarios. For their target trajectories $P_1(t)$ and $P_2(t)$, $\sum_{i=1}^2(\bar P-P_i)^2=2[\bar P-(P_1+P_2)/2]^2+(P_1-P_2)^2/2$. Dropping the first term, substituting the workload-dependent targets, and integrating gives~\eqref{eq:perc1_lower}. \hfill$\square$\par}

The UPS connection law $\gamma_h(V(t))\,r_{\mathrm{nom}}(t;\tau_{\mathrm{ramp}})$ defined by~Eq.~\eqref{eq:ups_gamma} and~Eq.~\eqref{eq:ramp} is one instance of $\Phi[V](t)$.

The bound is exact for the IT branch. Over an electrical-transient interval $T_{\mathrm{e}}\ll\tau_{\mathrm{th}}$, it extends approximately to the facility response when the thermal state and the initial motor and protection states are common to the two scenarios; otherwise, it should be interpreted as an IT-branch result. 

\vspace{-0.2cm}
\subsection{Grey-Box Architecture}
\label{sec:framework_arch}

The estimator receives the length-$L$ input window $\bm{x}_{t-L+1:t}$, sampled at the estimator resolution
$\Delta t_{\mathrm{eff}}=10$~ms. The output at time $t$ is the aggregate PCC vector together with the per-branch reconstructions,
\begin{equation}
\hat{\bm{y}}_t = \big[\hat P_{\mathrm{DC}}, \hat Q_{\mathrm{DC}}\big]^\top,\quad
\{\hat P_k, \hat Q_k\}_{k\in\{\mathrm{IT,cool,aux}\}}.
\label{eq:output_tensor}
\end{equation}
The estimator combines three components.

\emph{(a) Temporal backbone.} A dual-branch $1$D convolutional encoder processes the input at the native resolution and at a further-downsampled $100$~ms resolution, capturing the fast electrical and slow workload timescales, respectively. The encoded features feed a two-layer gated recurrent unit (GRU), producing a temporal representation $\bm h_t$ for the reconnect-refinement module.

\emph{(b) Physics-structured decomposition head.} The head instantiates the branch equations of Section~\ref{sec:model} with the learned workload law $g_{\mathrm{edge}}(w)$ and estimated cooling share $\hat\rho$ substituted for $g(w)$ and $\rho$, and with the reactive counterparts following the same structural expressions. In the cooling head, the motor slip is advanced at $\Delta t_{\mathrm{eff}}=10$~ms by the explicit Euler discretization of Eq.~\eqref{eq:slip},
\begin{equation}
s_{t+1} = s_t + \frac{\Delta t_{\mathrm{eff}}}{2H}\bigl[T_{\mathrm{mech}}(\hat T_{\mathrm{room},t}) - T_{\mathrm{em}}(V_t,s_t)\bigr].
\label{eq:slip_discrete}
\end{equation}
The motor slip $s_{t-L+1}$ that initializes each window is inherited from a deterministic per-scenario rollout of~Eq.~\eqref{eq:slip_discrete} launched from the pre-fault equilibrium $s_0$, so simulator-supplied slip is not used as an estimator input. At each sample the head evaluates $\bar P_{\mathrm{IM}}(V_t,s_t)$ and $\bar Q_{\mathrm{IM}}(V_t,s_t)$, forms the cooling-branch output, and then advances the state.

The UPS submodel uses the estimator gate
\begin{align}
\hat\gamma_{\mathrm{est}}(V,t) &= \gamma_h(V)\,\hat r_{\mathrm{est}}(t),\\
\hat r_{\mathrm{est}}(t) &= \operatorname{clip}\!\bigl[r_{\mathrm{nom}}(t;\tau_{\mathrm{ramp}}) \notag\\
&\qquad + \mathbf{1}\{t\ge t_{\mathrm{re}}\}\,\delta r_\theta(\tau_{\mathrm{since\,re}},\bm{h}_t),\,0,1\bigr],
\label{eq:gamma_est}
\end{align}
where $\gamma_h(V)$ is inherited from~Eq.~\eqref{eq:ups_gamma}, $r_{\mathrm{nom}}$ is the nominal linear ramp of~Eq.~\eqref{eq:ramp} at the anchored $\tau_{\mathrm{ramp}}$, and $\delta r_\theta$ is a small MLP residual conditioned on $(\tau_{\mathrm{since\,re}},\bm{h}_t)$ that absorbs reconnect-shape variation beyond the nominal ramp. The auxiliary feature $\tau_{\mathrm{since\,re}}$ is set to zero before the reconnect-eligibility event (the up-crossing of $V_{\mathrm{th}}+\Delta V$ in~Eq.~\eqref{eq:t_re}) and equals $t-t_{\mathrm{re}}$ thereafter; it is supplied only to this module. The cooling head evaluates $\Pi_P,\Pi_Q$ from $(V_t,s_t,\hat\rho,\hat T_{\mathrm{room},t})$ alone.

\emph{(c) Learned calibration elements.} The estimator learns the workload law $g_{\mathrm{edge}}(w)$ and the
cooling share $\hat\rho$. The workload law follows the univariate KAN construction of Liu \etal{}~\cite{liu2025kan},
\begin{equation}
g_{\mathrm{edge}}(w) = \sum_{j} c_j\,B_j(w) + b(w),
\label{eq:kan_edge}
\end{equation}
where $\{B_j\}$ is a cubic B-spline basis on a uniform knot grid over $w\in[0,1]$, $\bm c=[c_j]_j$ are trainable spline coefficients, and $b(w)$ is a trainable scaled-SiLU base term. Let $\bm\theta_g$ collect all trainable parameters of $g_{\mathrm{edge}}$.   The Supercloud-anchored reference law used to generate the branch-level training targets satisfies $g(w_{\mathrm{ref}})=1$
(Section~\ref{sec:case_interface}), but the KAN parameterization does not impose
$g_{\mathrm{edge}}(w_{\mathrm{ref}})=1$ as a hard constraint. The DOL/VFD share is represented as $\hat\rho=\sigma(\rho_{\mathrm{logit}})$ with $\rho_{\mathrm{logit}}\in\mathbb{R}$ a scalar parameter and $\sigma(\cdot)$ the sigmoid.

\vspace{-0.3cm}
\subsection{Two-Phase Training}
\label{sec:framework_train}

Training proceeds in two phases with a composite objective
\begin{equation}
\mathcal{L} = \mathcal{L}_{\mathrm{data}} + \lambda_{\mathrm{d}}\mathcal{L}_{\mathrm{decomp}} + \lambda_{\mathrm{s}}\mathcal{L}_{\mathrm{sparse}}.
\label{eq:loss}
\end{equation}
The aggregate-fidelity term $\mathcal{L}_{\mathrm{data}}$ is the mean-squared error (MSE) on the PCC power vector
\begin{equation}
\mathcal{L}_{\mathrm{data}}=\frac{1}{|\mathcal{B}|}\sum_{t\in\mathcal{B}}\Big[(\hat P_{\mathrm{DC}}(t){-}P_{\mathrm{DC}}(t))^2+(\hat Q_{\mathrm{DC}}(t){-}Q_{\mathrm{DC}}(t))^2\Big],
\label{eq:loss_data}
\end{equation}
where $\mathcal{B}$ indexes the samples in a mini-batch. The decomposition term $\mathcal{L}_{\mathrm{decomp}}$ is the MSE over branches $k\in\{\mathrm{IT},\mathrm{cool},\mathrm{aux}\}$
\begin{equation}
\mathcal{L}_{\mathrm{decomp}} = \sum_{k}\frac{1}{|\mathcal{B}|}\sum_{t\in\mathcal{B}}\!\big[(\hat P_{k,t}{-}P_{k,t})^2 + (\hat Q_{k,t}{-}Q_{k,t})^2\big],
\label{eq:loss_decomp}
\end{equation}
and $\mathcal{L}_{\mathrm{sparse}}=\Vert\bm{c}\Vert_1$ is an $\ell_1$ penalty on the KAN spline coefficients that encourages a sparse spline representation. 

\emph{Phase~1 (joint training).} The temporal backbone, the reconnect-refinement MLP $\delta r_\theta$, the workload-edge parameters $\bm\theta_g$, and $\rho_{\mathrm{logit}}$ are optimized jointly with the Adam optimizer on the composite objective~Eq.~\eqref{eq:loss}.

\emph{Phase~2 (targeted refinement).} 
The temporal backbone and the reconnect-refinement MLP are frozen, only
$\bm\theta_g$ and $\rho_{\mathrm{logit}}$ are updated. The motor
parameters, including $H$ and $T_{\mathrm{m},0}$, remain fixed throughout
training. 

\subsection{Symbolic Distillation and Deployment}
\label{sec:framework_distill}

The KAN edge $g_{\mathrm{edge}}$ resulting from Phase~2 is expressed in a cubic B-spline basis. While accurate, it is less convenient for direct implementation in a transient-stability tool. Post-training symbolic distillation extracts a closed-form replacement. The trained edge is evaluated on a uniform grid of $n$ points $\{w_i\}_{i=1}^{n}\subset[0,1]$, producing samples $\{\hat g_i\}=\{g_{\mathrm{edge}}(w_i)\}$.
A compact library of six one-dimensional parametric families is fit to these samples by nonlinear least squares, spanning linear, quadratic, cubic, four-parameter sigmoid, saturating-exponential, and offset power-law forms $a_0 + a_1 w^{a_2}$.
For each candidate family $\mathcal{H}_j$ with $k_j$ parameters and residual sum of squares $\mathrm{RSS}_j = \sum_i (\hat g_i - g_j(w_i))^2$, the Bayesian Information Criterion is
\begin{equation}
\mathrm{BIC}_j = n\log(\mathrm{RSS}_j/n) + k_j\log n,
\label{eq:bic}
\end{equation}
and the family minimizing $\mathrm{BIC}_j$ yields the closed form
$g_{\mathrm{sym}}(w)$. Within the estimator, $g_{\mathrm{sym}}(w)$
replaces $g_{\mathrm{edge}}(w)$, while $\delta r_\theta$ remains part of
$\hat r_{\mathrm{est}}$ in~Eq.~\eqref{eq:gamma_est}.

For solver deployment (Section~\ref{sec:case_closedloop}), the composite
model uses $g_{\mathrm{sym}}(w)$, the recovered share $\hat\rho$, and the
prescribed gate
$\gamma_h(V)\,r_{\mathrm{nom}}(t;\tau_{\mathrm{ramp}})$.
The settings $V_{\mathrm{th}}$ and $\Delta V$ are prescribed, while
$\tau_{\mathrm{ramp}}$ is externally anchored. The remaining model
parameters are facility-specific engineering inputs. The estimator-only
residual $\delta r_\theta$ is not deployed in the solver.
\vspace{-0.5em}
\section{Case Studies}
\label{sec:case}

This section first examines workload-law anchoring and recovery of the
selected calibration elements. We then evaluate predictive accuracy and
use ablations to isolate the roles of workload telemetry and branch-level
supervision. Finally, we test the recovered workload law and cooling
composition across independently implemented simulator variants, examine
generalization to unseen voltage sags, and evaluate the closed-loop system
response.
\vspace{-0.5em}
\subsection{Experimental Setup}
\label{sec:case_setup}

The evaluation combines a workload law anchored to operational GPU
telemetry with simulated grid responses. The workload-law anchor is fitted
to $300{,}000$ per-GPU power and utilization samples from the MIT
Supercloud NVIDIA-SMI telemetry~\cite{samsi2021supercloud}, recorded at
approximately $100$-ms intervals. Grid responses are simulated on the
IEEE~39-bus system using ANDES~1.10.1, with the data-center facility
connected at bus~15. The dataset covers bus-fault and line-trip events
over a range of voltage-sag severities, three facility ratings
($100$, $300$, and $500$~MW), and three workload patterns (constant,
batched, and AI pulsing), yielding $12{,}600$ scenarios. Before sliding-window extraction, the scenarios are divided using an
event-grouped split based on disturbance identity. The $69$ distinct
$(\text{fault type},V_{\mathrm{sag}})$ identities are assigned $47/10/12$
to the training, validation, and test sets, respectively, so that all
scenarios and windows sharing a disturbance identity remain in the same
partition.

{\color{black}The reference facility uses $(\alpha_{\mathrm{IT}},\alpha_{\mathrm{cool}},\alpha_{\mathrm{aux}})=(0.55,0.35,0.10)$, $V_{\mathrm{th}}=0.85$~p.u., $\Delta V=0.05$~p.u., and $\tau_{\mathrm{ramp}}=2.0$~s~\cite{jimenez2025data}. The VFD dropout threshold is $V_{\mathrm{d,VFD}}=0.90$~p.u.; the cooling relay uses $(V_{\mathrm{relay}},\tau_{\mathrm{dwell}},\tau_{\mathrm{delay}})=(0.50~\mathrm{p.u.},150~\mathrm{ms},50~\mathrm{ms})$, and $\tau_{\mathrm{th}}=120$~s. The simulator and estimator use the same fixed reference motor parameters.}

Five terminal-response baselines are used: two neural models, a feedforward ANN and {\color{black}an LSTM}, and three parametric
composites, namely a ZIP model with a fixed-slip induction motor
(ZIP+IM), the WECC composite-load aggregate~\cite{wecc_clm}, and an
independent reimplementation of the Jimenez data-center
load model~\cite{jimenez2025data}. For the Jimenez model, the published
motor-equivalent-circuit and voltage-gate constants are held fixed,
leaving five free parameters for calibration. The neural baselines use
the same optimizer, learning rate, batch size, and epoch budget as
Phase~1 of the proposed framework. The ANN and LSTM use the same
terminal input channels but are trained on aggregate $P/Q$ targets
only; the proposed estimator also uses branch-level $P/Q$ targets. Each parametric baseline is calibrated
separately for each workload pattern, while the proposed estimator uses
a single parameterization across all workload patterns. {\color{black}Detailed training and baseline configurations are provided
in the online repository.}

Throughout this section, ``transient'' metrics are computed over samples
with $V(t)\leq0.9$~p.u.\ and the $2$-s recovery interval after the voltage
subsequently returns above $0.9$~p.u.; ``overall'' metrics use all samples.
Each metric is computed per test window and then averaged across the test
windows. Unless otherwise stated, all training and retraining experiments
for the proposed estimator and neural baselines use three random seeds.
Reported learning-based metrics are seed means, and uncertainty values
denote one standard deviation across seeds.
\vspace{-0.5cm}
\subsection{Anchoring and Recovery of Calibration Elements}
\label{sec:case_interface}

\emph{1) Workload-law anchoring and reference-case recovery.}
Fig.~\ref{fig:workload_law} shows the per-GPU telemetry samples and
binwise means used to anchor $g(w)$. The binwise means show a nonlinear
power--utilization relationship and yield an idle-power ratio of $0.272$.
After subtracting the idle component, we normalize the remaining power
increment by its full-utilization value and fit the constrained cubic
$h(w)=0.5721w+0.7496w^2-0.3217w^3$, with $h(0)=0$ and $h(1)=1$. The anchored workload law is
\begin{align}
g_{\mathrm{ref}}(w)
&=0.272+(1-0.272)h(w) \notag\\
&=0.272+0.4165w+0.5457w^2-0.2342w^3 .
\label{eq:gofw_cubic}
\end{align}
The fit achieves $R^2=0.976$ on the binwise means and is used in the
composite simulator to generate the training and evaluation responses.
The estimator uses the simulator's branch structure and anchored motor
parameters, so the following calibration results assess recoverability
under a correctly specified physical structure using simulated PCC
responses.
{
\setlength{\abovecaptionskip}{-2pt}  
\setlength{\belowcaptionskip}{-10pt} 
\begin{figure}[t]
\centering
\includegraphics[width=\linewidth]{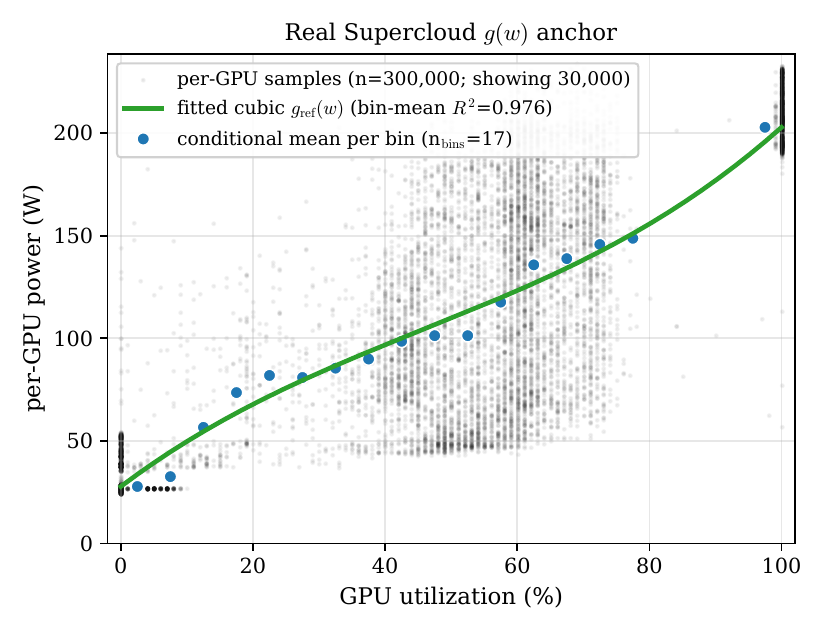}
\caption{MIT Supercloud workload-law anchor: $300{,}000$ per-GPU power and utilization samples, binned conditional means, and the fitted cubic.}
\label{fig:workload_law}
\end{figure}}

Table~\ref{tab:accuracy_3seed} reports branch-level test accuracy and
recovery statistics for the learned calibration elements. The learned KAN edge
$g_{\mathrm{edge}}(w)$ is distilled into a cubic law
$g_{\mathrm{sym}}(w)$, with a maximum relative coefficient error of
$7\%$. The distilled law achieves $R^2=0.9999$
against $g_{\mathrm{ref}}(w)$ over $w\in[0,1]$. At the reference setting
$\rho=0.50$, the recovered DOL/VFD share is
$\hat\rho=0.539\pm0.015$. The corresponding IT and cooling branch
predictions attain overall $R^2$ values of $0.957$ and $0.797$.

\begin{table}[t]
\centering
\caption{Held-out branch-prediction accuracy and calibration-element
recovery statistics.}
\label{tab:accuracy_3seed}
\scriptsize
\setlength{\tabcolsep}{3pt}
\begin{tabular}{@{}lcccc@{}}
\toprule
Element / branch & $R^2$ ovr. & $R^2$ trans. & RMSE ovr. & RMSE trans. \\
\midrule
$P_{\mathrm{IT}}$ branch    & $0.957$ & $0.975$ & $0.029$ & $0.020$ \\
$P_{\mathrm{cool}}$ branch  & $0.797$ & $0.866$ & $0.013$ & $0.027$ \\
\midrule
$g_{\mathrm{sym}}(w)$ vs.\ anchored reference & \multicolumn{4}{l}{$R^2=0.9999$, $\|g_{\mathrm{sym}}-g_{\mathrm{ref}}\|_\infty = 0.003$} \\
$\hat\rho$ (truth $0.50$) & \multicolumn{4}{l}{$0.539\pm 0.015$} \\
\bottomrule
\end{tabular}
\end{table}

\emph{2) Cooling-composition and reconnect-ramp duration.}
{\color{black}The nuisance-projected information for $\rho$ is evaluated at $\rho_0=0.5$ with $\bm\eta=(H,T_{\mathrm{m},0},s_0,R_s,R_r,X_s+X_r,X_m,q_{\mathrm{disp}})$, $\sigma_P=\sigma_Q=0.005$~facility p.u., $10$-ms sampling, and a $1.2$-s window from fault inception. These eight quantities are profiled in the FIM calculation, although the motor parameters remain fixed during estimator training. The cooling branch is kept connected and its scale is anchored, consistent with Theorem~\ref{thm:rho}. For the $300$-ms sags with $V_{\min}<0.5$~p.u., this is a prescribed connected-branch excitation with cooling-relay operation suppressed.}

Across the excitation grid, the numerical rank is $6$ for $J_\eta$ and
$7$ for $[J_\eta\ \bm j_\rho]$. The added $\rho$ sensitivity therefore
remains outside the numerical column space of the nuisance Jacobian. Under
the assumed noise model, the Cram\'er--Rao bound (CRB) gives the
standard-deviation lower bound
$\sigma_{\rho,\mathrm{CRB}}
=\mathcal{I}_{\rho\mid\eta}^{-1/2}$. For the $300$-ms sags,
$\sigma_{\rho,\mathrm{CRB}}$ decreases from
$9.0\times10^{-3}$ at $V_{\min}=0.85$ to
$5.8\times10^{-3}$ at $V_{\min}=0.17$. At the same
$V_{\min}=0.17$, shortening the sag to $100$~ms increases the bound to
$1.72\times10^{-2}$. Under the connected-branch setting, sag depth and duration
jointly determine the information remaining after nuisance projection.
When an unknown common cooling-branch gain is added to $\bm\eta$,
$\mathcal{I}_{\rho\mid\eta}$ becomes numerically negligible, showing the
dependence on the anchored cooling scale.
{\color{black}The full excitation sweep is reported in the online repository.}

Fig.~\ref{fig:recoverability}(a) reports a separate fixed-nuisance
recovery sweep. The estimator is retrained at
$\rho_{\mathrm{true}}\in\{0.30,0.40,0.50,0.60,0.70\}$, with all other
physical parameters held at their reference values. The mean
estimate increases monotonically with $\rho_{\mathrm{true}}$ with a
fitted slope of $0.89$. The estimates exceed the true values by
approximately $0.05$--$0.10$, and the standard deviation is no greater
than $0.023$. The estimator therefore tracks the cooling composition
with a positive offset under these fixed-nuisance conditions. Together
with the FIM result, this sweep supports estimating $\rho$ while
retaining the cooling scale as external information.

For the reconnect-ramp duration, the architecture does not expose
$\tau_{\mathrm{ramp}}$ as a scalar estimate. An equivalent ramp duration
$\tau_{\mathrm{eq}}$ is reconstructed by sweeping
$\tau_{\mathrm{since\,re}}$ through the reconnect-refinement MLP while
fixing the temporal representation at the reconnect-eligibility event.
Across $453$ scenarios with imposed reconnect-ramp durations
$\tau_{\mathrm{true}}\in[1.0,3.0]$~s, $\tau_{\mathrm{eq}}$ remains near
$1$~s and has little correlation with $\tau_{\mathrm{true}}$
(Pearson $r=0.018$, Fig.~\ref{fig:recoverability}(b)). The residual therefore does not yield a scalar
timescale that tracks the imposed ramp duration in this diagnostic.
Accordingly, $\tau_{\mathrm{ramp}}$ remains externally specified for
deployment.
{
\setlength{\abovecaptionskip}{-2pt}  
\setlength{\belowcaptionskip}{-12pt} 
\begin{figure}[t]
\centering
\includegraphics[width=\linewidth]{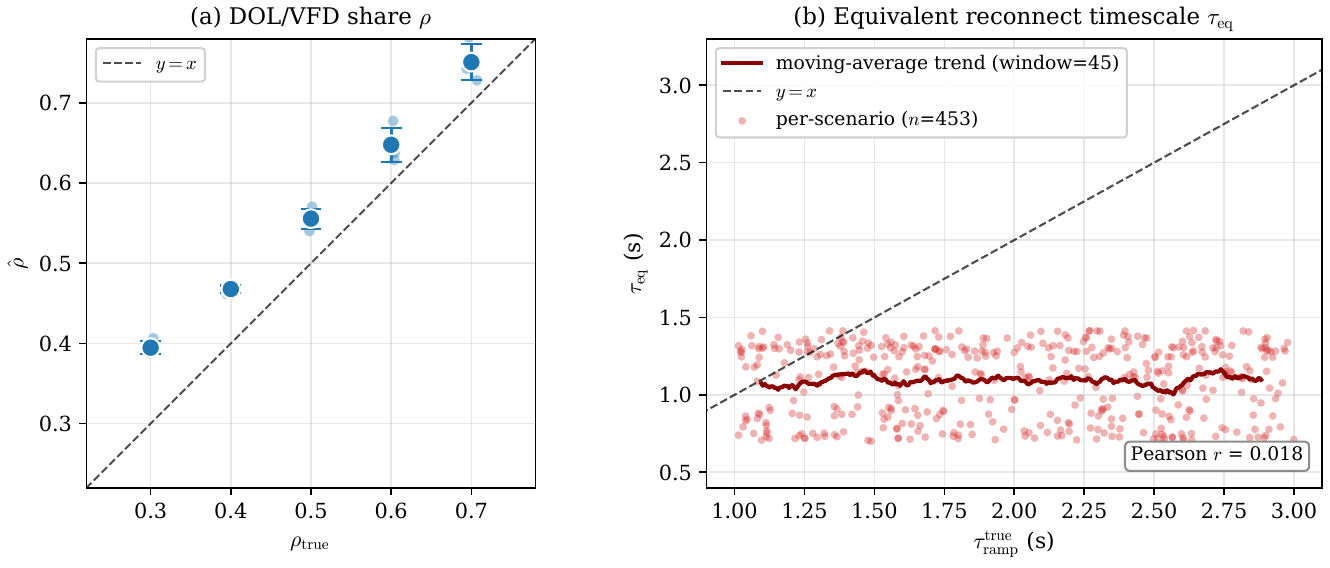}
\caption{Empirical recoverability check. (a) Learned $\hat\rho$ vs ground truth $\rho_{\mathrm{true}}$ in a fixed-nuisance controlled sweep. (b) Equivalent reconnect-ramp duration $\tau_{\mathrm{eq}}$. }
\label{fig:recoverability}
\end{figure}}

Varying the estimator's assumed reconnect threshold $V_{\mathrm{th}}+\Delta V$
from $0.85$ to $0.95$~p.u., around the simulator setting of $0.90$~p.u.,
changes transient $P_{\mathrm{IT}}$ $R^2$ by $0.010$. This variation is of the
same order as the standard deviation of $0.005$ at the
reference setting. {\color{black}The tested threshold settings are listed in the online repository.}

\subsection{Predictive Accuracy and Ablation}
\label{sec:case_accuracy}

\emph{1) Baseline comparison.}
Table~\ref{tab:baselines} reports overall $R^2$ and overall and transient
RMSE for $P$ and $Q$ on the event-grouped test set. 
The proposed estimator reduces overall $P$ RMSE from $0.057$ for the
LSTM to $0.035$, a $39\%$ reduction, while using $3.3\times$ fewer
trainable parameters. The LSTM has lower transient $P$ RMSE
($0.028$ versus $0.031$) and lower transient $Q$ RMSE. The two models
have similar overall $Q$ accuracy. The proposed estimator additionally
provides branch-level reconstructions and a distilled closed-form
workload law.

The feedforward ANN has higher transient $P$ error than both recurrent
models. ZIP+IM and the WECC composite-load aggregate give nearly
identical overall $P$ RMSE values of $0.145$. The calibrated Jimenez
model, which uses a DOL-only cooling representation, reduces this error
to $0.071$ but shows a $Q$ bias of $+0.05$~p.u.

Fig.~\ref{fig:trajectory_indist} shows a grouped-test case with
$V_{\min}=0.77$. Panel~(b) compares the predicted
$P_{\mathrm{IT}}$ and $P_{\mathrm{cool}}$ trajectories with their
simulator targets. Both branch predictions follow the corresponding
targets, including the post-fault cooling recovery, rather than only
matching their aggregate sum.

\begin{table*}[t]
\centering
\caption{Baseline comparison on the event-grouped test set. }
\label{tab:baselines}
\footnotesize
\setlength{\tabcolsep}{3pt}
\begin{tabular}{@{}lc ccc ccc c@{}}
\toprule
& & \multicolumn{3}{c}{$P$} & \multicolumn{3}{c}{$Q$} & \\
\cmidrule(lr){3-5}\cmidrule(lr){6-8}
Baseline & \#params & overall $R^2$ & ovr.\ RMSE & trans.\ RMSE & overall $R^2$ & ovr.\ RMSE & trans.\ RMSE & Workload-law form \\
\midrule
ANN & $17{,}410$ & $0.430$ & $0.1137$ & $0.1409$ & $0.761$ & $0.0070$ & $0.0132$ & none \\
LSTM  & $51{,}330$ & $0.856$ & $0.0571$ & $\mathbf{0.0278}$ & $0.953$ & $0.0031$ & $\mathbf{0.0040}$ & none \\
ZIP + IM & $7$ & $0.075$ & $0.1453$ & $0.1728$ & $0.510$ & $0.0101$ & $0.0154$ & none \\
WECC CLM aggregate & $11$ & $0.079$ & $0.1450$ & $0.1765$ & $0.515$ & $0.0100$ & $0.0150$ & none \\
Jimenez--calibrated & $5$ & $0.780$ & $0.0708$ & $0.1075$ & $-13.765$ & $0.0553$ & $0.0410$ & fixed \\
\textbf{Proposed} & $15{,}464$ & $\mathbf{0.948}$ & $\mathbf{0.0346}$ & $0.0307$ & $\mathbf{0.959}$ & $\mathbf{0.0029}$ & $0.0080$ & distilled \\
\bottomrule
\end{tabular}
\end{table*}

{
\setlength{\abovecaptionskip}{-2pt}  
\setlength{\belowcaptionskip}{-15pt} 
\begin{figure}[t]
\centering
\includegraphics[width=\linewidth]{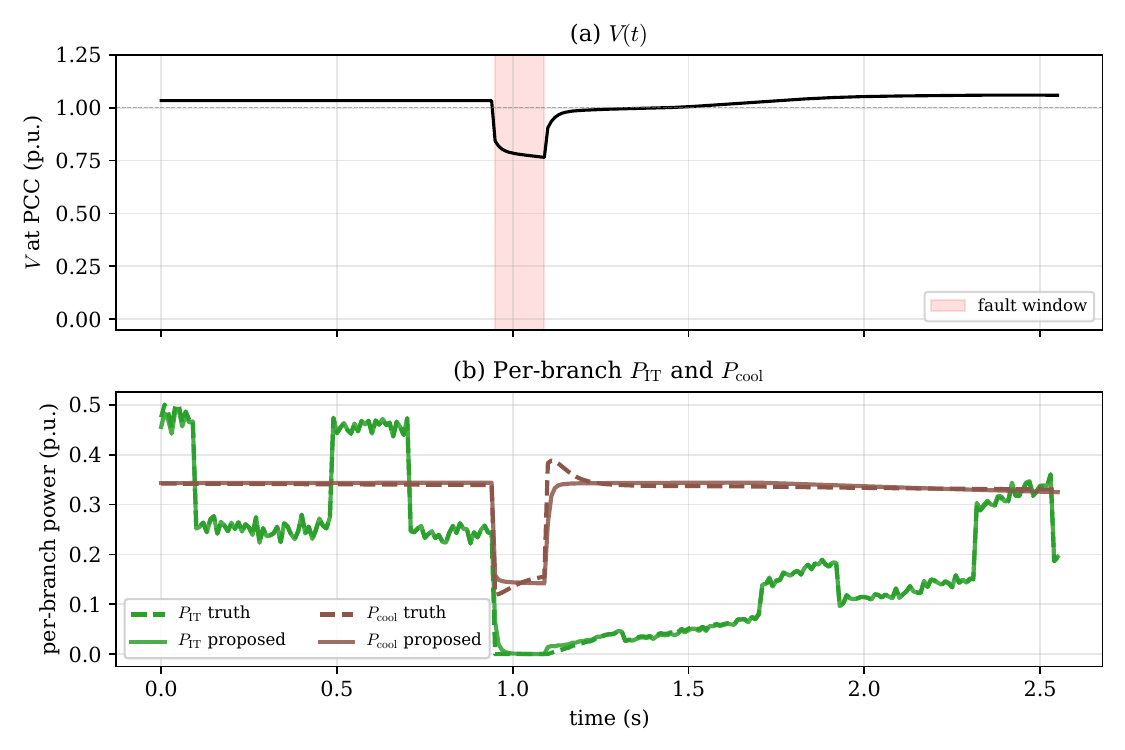}
\caption{Representative grouped-test trajectory ($V_{\min}=0.77$). (a)~PCC voltage. (b)~Per-branch $P_{\mathrm{IT}}$ and $P_{\mathrm{cool}}$.}
\label{fig:trajectory_indist}
\end{figure}}

\emph{2) Ablation study.}
Two ablations assess the roles of workload telemetry and branch-level
supervision. In the first, $w(t)$ is masked to zero at both the encoder
input and the physics-head evaluation of $g_{\mathrm{edge}}(w)$. In the
second, $\mathcal{L}_{\mathrm{decomp}}$ is removed so that training uses
only the aggregate $P/Q$ objective. Both variants are retrained on the
same event-grouped training set. Table~\ref{tab:ablation} reports the
grouped-test results.

Masking $w(t)$ reduces aggregate $P$ $R^2$ from $0.948$ to $0.582$ and
IT-branch $R^2$ from $0.957$ to $0.443$. The cooling-branch transient
$R^2$ changes only from $0.866$ to $0.830$. This selective degradation
is consistent with the factorization
$P_{\mathrm{IT}}\propto g(w)$ in~Eq.~\eqref{eq:p_it} and with the
workload-input requirement in Proposition~\ref{prop:cmpldw}.

Removing $\mathcal{L}_{\mathrm{decomp}}$ reduces aggregate $P$ $R^2$
only from $0.948$ to $0.924$, but the cooling-branch transient $R^2$
falls from $0.866$ to $-0.435$ and $\hat\rho$ shifts from $0.539$ to
$0.189$. The aggregate objective constrains the branch sum but not its
allocation, allowing the cooling response and composition to drift while
the terminal response remains accurate. Aggregate trajectory accuracy
therefore does not establish recovery of the intended decomposition.

Together, these results show that workload telemetry is required to
represent workload-dependent IT demand, while supervision with the full
set of branch $P/Q$ targets prevents an accurate aggregate fit from
masking an incorrect cooling decomposition.

\begin{table}[t]
\centering
\caption{Workload-channel and decomposition-supervision ablation on the event-grouped test set. }
\label{tab:ablation}
\scriptsize
\setlength{\tabcolsep}{3pt}
\begin{tabular}{@{}lccccc@{}}
\toprule
Configuration & $P$ RMSE & $P$ $R^2$ & $P_{\mathrm{IT}}$ $R^2$ & $P_{\mathrm{cool}}$ $R^2$ tr. & $\hat\rho$ \\
\midrule
Full (reference)      & $0.034$ & $0.948$ & $0.957$ & $0.866$        & $0.539$ \\
$-$ workload input $w$      & $0.098$ & $0.582$ & $0.443$ & $0.830$        & $0.531$ \\
$-$ $\mathcal{L}_{\mathrm{decomp}}$ (aggr.-only)   & $0.042$ & $0.924$ & $0.964$ & $-0.435$ & $0.189$ \\
\bottomrule
\end{tabular}
\end{table}

\vspace{-0.5em}
\subsection{Robustness and Generalization}
\label{sec:case_generalization}

\emph{1) Cross-implementation evaluation.}
We retrain the estimator on two variants generated by an independently
implemented simulator to test whether recovery of the learned calibration
elements depends on the reference implementation. The variants change the workload-law idle
ratio and cubic coefficients, the DOL/VFD share $\rho$, and
$\tau_{\mathrm{ramp}}$, while retaining the reference motor parameters.
The motor state is initialized from the corresponding pre-fault torque
equilibrium in~Eq.~\eqref{eq:slip_equilibrium}. During retraining,
$\tau_{\mathrm{ramp}}$ remains anchored at $2.0$~s, leaving
$\delta r_\theta$ to represent the variant-specific reconnect shape.
The event-grouped split is applied separately to each variant. {\color{black}The full variant settings and scenario counts are documented in the online repository.}

The two variants have distinct workload-law shapes, with the coefficient
of $w^3$ reversing sign between them. As shown in
Table~\ref{tab:cross_impl}, the proposed estimator achieves overall
$P$ $R^2$ values of $0.923$ and $0.915$ on Variants~1 and~2,
respectively. The recovered workload edge has
$R^2\geq0.999$ against the corresponding variant law, and the distilled
cubic coefficients have maximum relative errors of $3.7\%$ and $1.9\%$.
The recovered $\hat\rho$ values differ from their true values by
$+0.04$ and $-0.05$ for Variants~1 and~2, respectively. On Variant~2,
the transient $R^2$ is $0.908\pm0.014$, compared with
$0.593\pm0.347$ for the retrained LSTM.
These results show that, after retraining, recovery of the workload law and
cooling-composition is maintained across the two tested
simulator implementations.

\begin{table}[t]
\centering
\caption{Cross-implementation retraining on two variants generated by the independently implemented simulator.}
\label{tab:cross_impl}
\scriptsize
\setlength{\tabcolsep}{2pt}
\begin{tabular}{@{}lccccc@{}}
\toprule
Variant &$P$ RMSE ovr. & RMSE tr. &$P$ $R^2$ ovr. & $g_{\mathrm{edge}}$ $R^2$ & $\hat\rho$ (truth) \\
\midrule
V1 & $.037{\pm}.003$ & $.044{\pm}.004$ & $.923{\pm}.013$ & $.9995{\pm}.0002$ & $0.44\,(0.40)$ \\
V2 & $.049{\pm}.004$ & $.046{\pm}.004$ & $.915{\pm}.012$ & $.9992{\pm}.0003$ & $0.55\,(0.60)$ \\
\bottomrule
\end{tabular}
\end{table}

\emph{2) Voltage-shift generalization.}
We next evaluate voltage conditions outside the training range. Training
is restricted to scenarios with $V_{\min}>0.75$~p.u., while testing uses
only deep-sag scenarios with $V_{\min}<0.65$~p.u. Scenarios with
$V_{\min}\in[0.65,0.75]$~p.u. are excluded from both sets to provide a
buffer between the training and test ranges. The resulting sets contain
$6{,}894$ training scenarios and $4{,}188$ deep-sag test windows.
Table~\ref{tab:ood} reports the aggregate test metrics, and
Fig.~\ref{fig:ood} shows a representative trajectory.

On the deep-sag test set, the proposed estimator retains a transient
$R^2$ of $0.750\pm0.014$, compared with $0.033\pm0.207$ for the LSTM
and $-1.43\pm1.03$ for the ANN. Its transient $R^2$ on the
in-distribution shallow-sag test set is $0.935\pm0.014$. The voltage
shift therefore reduces predictive accuracy, but the proposed estimator
retains substantially higher and more consistent transient accuracy than
the neural baselines. With the branch structure and motor parameters
fixed across the voltage split, this result characterizes physics-guided
extrapolation to deeper sags.

\begin{table}[t]
\centering
\caption{Out-of-distribution voltage-generalization results.}
\label{tab:ood}
\scriptsize
\setlength{\tabcolsep}{2pt}
\begin{tabular}{@{}lcccc@{}}
\toprule
Model &$P$ RMSE ovr. & $P$ $R^2$ ovr. & $R^2$ tr. &$R^2$ tr.\ range \\
\midrule
Proposed & $\bm{0.050{\pm}0.001}$ & $\bm{0.941{\pm}0.003}$ & $\bm{0.750{\pm}0.014}$ & $[0.74, 0.77]$ \\
LSTM     & $0.105{\pm}0.008$      & $0.744{\pm}0.039$      & $0.033{\pm}0.207$     & $[-0.21, +0.30]$ \\
ANN      & $0.153{\pm}0.024$      & $0.441{\pm}0.194$      & $-1.43{\pm}1.03$      & $[-2.88, -0.58]$ \\
\bottomrule
\end{tabular}
\end{table}

{
\setlength{\abovecaptionskip}{-2pt}  
\setlength{\belowcaptionskip}{-15pt} 
\begin{figure}[t]
\centering
\includegraphics[width=\linewidth]{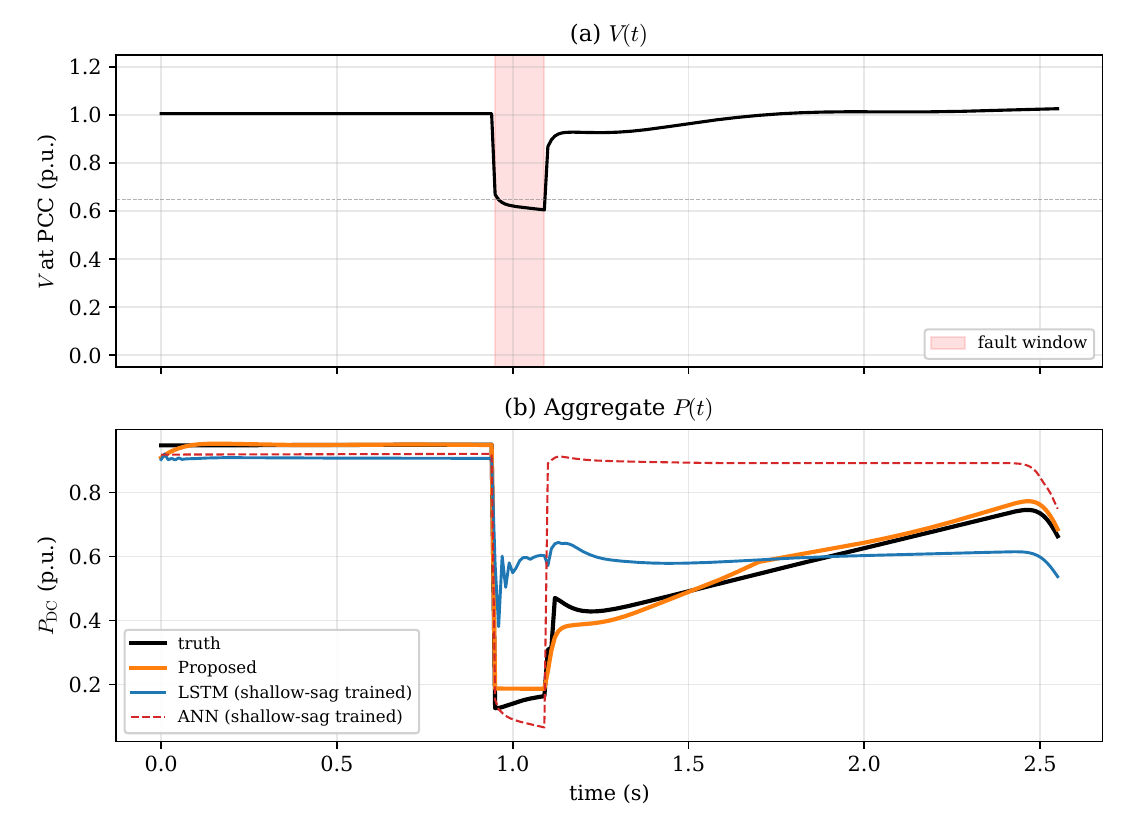}
\caption{OOD deep-sag test scenario ($V_{\min}=0.61$). (a)~PCC voltage. (b)~Aggregate $P_\mathrm{DC}(t)$.}
\label{fig:ood}
\end{figure}}

\vspace{-1em}
\subsection{Closed-Loop System-Level Evaluation}
\label{sec:case_closedloop}

Closed-loop behavior is evaluated by comparing
the proposed composite with a data-center-parameterized WECC~PERC1
aggregate. The fault matrix combines three facility sizes
($200$, $300$, and $500$~MW), three fault durations
($100$, $200$, and $300$~ms), and three fault locations
(bus~15, adjacent bus~16, and two-hop bus~4), giving $27$ fault
conditions. All faults are three-phase faults with
fault reactance $x_f=0.001$~p.u. Each condition is simulated using an open-loop PQ
representation, the proposed composite with recovered $\hat\rho$, the
PERC1 aggregate, and an endpoint version of the proposed composite with
$\rho=1$. The open-loop case provides the voltage prescan used to
parameterize PERC1.

The proposed composite uses the deployment model specified in
Section~\ref{sec:framework_distill}. Motor slip is initialized from
Eq.~\eqref{eq:slip_equilibrium} and advanced by the ANDES transient
solver using the anchored motor parameters.

Both representations use an IT/UPS cease threshold of $0.85$~p.u. For
each fault condition, the PERC1 cease and reconnect fractions are
parameterized from the corresponding pre-scanned open-loop
$V_{\min}$~\cite{kosterev2026perc1,weber2026perc1}. The proposed
composite instead determines cooling protection from the closed-loop
voltage trajectory. Its VFD fraction $(1-\hat\rho)$ drops out below
$0.90$~p.u.\ and re-engages upon voltage recovery. The cooling branch
trips only when $V<V_{\mathrm{relay}}=0.5$~p.u.\ persists for
$\tau_{\mathrm{dwell}}=150$~ms, followed by a
$\tau_{\mathrm{delay}}=50$~ms mechanical delay. {\color{black}PERC1 uses $F_{\mathrm{cease}}=0.55$ for $V_{\min}\in[0.5,0.85]$~p.u. and $0.90$ for $V_{\min}<0.5$~p.u.; its deep-sag reconnect fraction is $0.65$. Its reconnect delay and ramp are $0.5$ and $3$~s, respectively, compared with no added reconnect delay and a $2$-s ramp for the proposed composite.}
{\color{black}The complete protection and reconnection settings are tabulated in the online repository.}

{
\setlength{\abovecaptionskip}{-2pt}  
\setlength{\belowcaptionskip}{-15pt} 
\begin{figure*}[t]
\centering
\includegraphics[width=\linewidth]{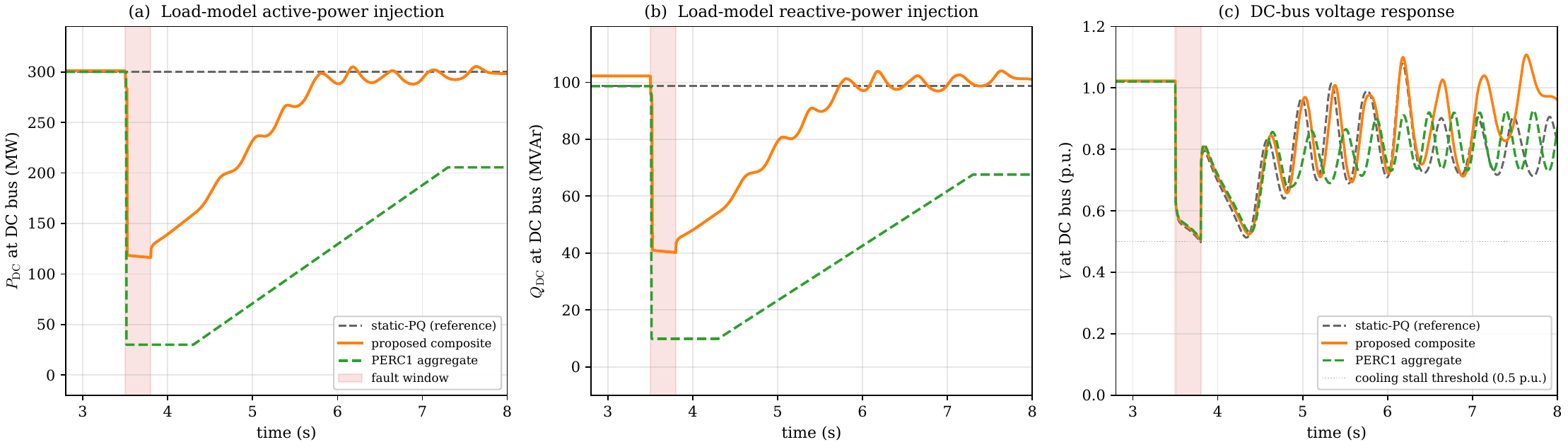}
\caption{Boundary-case closed-loop response for a $300$-MW
facility under a $300$-ms fault at two-hop bus~4.
\textbf{(a)} Active- and \textbf{(b)} reactive-power injections
at the DC bus. \textbf{(c)} DC-bus voltage.}
\label{fig:pathA_p_dc}
\end{figure*}}

For each pair of load-model configurations $(a,b)$, the maximum voltage discrepancy at the DC bus is
\begin{equation}
\Delta V_{\max}^{(a,b)} = \max_{t\in[t_{\mathrm{fault}},\,t_{\mathrm{fault}}+4.5\,\mathrm{s}]}\bigl|V^{(a)}(t) - V^{(b)}(t)\bigr|.
\end{equation}
Table~\ref{tab:closed_loop_gap} reports $\Delta V_{\max}$ aggregated across the $27$ unique fault configurations and stratified by duration and location. Dashes indicate percentiles not reported for the small stratified subsets.

\begin{table}[t]
\centering
\caption{Closed-loop DC-bus voltage discrepancies across the $27$ fault conditions.}
\label{tab:closed_loop_gap}
\footnotesize
\setlength{\tabcolsep}{4pt}
\begin{tabular}{@{}lrrrr@{}}
\toprule
Comparison & median & $p75$ & $p90$ & max \\
\midrule
Open vs.\ proposed composite            & 0.028 & 0.044 & 0.178 & 0.371 \\
Open vs.\ PERC1 aggregate            & 0.029 & 0.049 & 0.166 & 0.377 \\
PERC1 aggregate vs.\ proposed composite        & 0.010 & 0.014 & 0.044 & 0.334 \\
Endpoint $\rho{=}1$ vs.\ recovered $\hat\rho$ composite & 0.005 & 0.008 & 0.088 & 0.364 \\
\midrule
\multicolumn{5}{l}{\textit{Stratified (PERC1 aggregate vs.\ proposed composite):}} \\
\quad 100~ms, all buses          & 0.010 & -- & -- & 0.030 \\
\quad 200~ms, all buses          & 0.006 & -- & -- & 0.012 \\
\quad 300~ms, bus~4 (2-hop)      & 0.212 & -- & -- & 0.334 \\
\quad 300~ms, bus~15 (DC bus)    & 0.006 & -- & -- & 0.010 \\
\quad 300~ms, bus~16 (adjacent)  & 0.008 & -- & -- & 0.010 \\
\bottomrule
\end{tabular}
\end{table}

The proposed composite and PERC1 agree closely in $24$ of the $27$ fault
conditions. Across all $100$- and $200$-ms faults and the $300$-ms faults at buses~15 and~16, the maximum DC-bus voltage discrepancy does not exceed $0.030$~p.u. The discrepancy increases only for the three $300$-ms faults at bus~4, where the DC-bus voltage trajectory approaches the cooling-protection boundary.

In these three scenarios, PERC1 selects its reconnect fraction from the pre-scanned open-loop $V_{\min}$, whereas the proposed composite retains $\chi_{\mathrm{cool}}=1$ because the excursion below $V_{\mathrm{relay}}$ does not persist for $\tau_{\mathrm{dwell}}$. The proposed composite therefore restores the full pre-fault load, while PERC1 retains approximately $68\%$, producing the upper tail of the voltage discrepancy shown in Fig.~\ref{fig:pathA_p_dc} ($300$~MW facility, $V_{\min}\approx 0.50$).

The endpoint sensitivity further shows why the recovered cooling composition matters in closed-loop simulation. Replacing $\hat\rho=0.539$ with the all-DOL endpoint $\rho=1$ changes little over most of the fault matrix, but produces voltage deviations of up to $0.36$~p.u.\ in the same three bus-4 faults. In this boundary regime, the DOL/VFD share alters the cooling-branch voltage feedback and shifts the closed-loop trajectory to a different post-fault operating point, so the recovered $\rho$ affects not only trajectory fit but also the retained-load state the closed-loop simulation reaches.

Together, these tests identify a narrow protection-boundary regime in
which both the choice of load-model implementation and the DOL/VFD share
can affect the predicted retained-load state and DC-bus
voltage.
\vspace{-0.8em}
\section{Conclusion}
\label{sec:conclusion}

This paper developed a branch-resolved data-center load model and an
{\color{black}identifiability-guided} framework that estimates selected calibration elements and distills the workload law for transient-stability
simulation. Tests on $12{,}600$ IEEE~39-bus fault scenarios showed that
the estimator reduced overall active-power RMSE by $39\%$ relative to an
LSTM and retained a transient $R^2$ of $0.750$ for voltage sags outside
the training range. More importantly, aggregate-only supervision
preserved the PCC response while producing an incorrect cooling response
and DOL/VFD share. Terminal accuracy alone is therefore not sufficient
for calibrating a model intended to retain physical meaning. In
closed-loop simulations, the proposed model and the PERC1 aggregate
produced similar voltage responses in $24$ of $27$ fault
conditions. The remaining cases occurred near the cooling-protection
boundary, where the two models reached different retained-load states.
Within the proposed model, changing only the DOL/VFD share to the
all-DOL endpoint produced voltage deviations of up to $0.36$~p.u.
These results show that cooling composition and the treatment of
protection and reconnection can affect closed-loop
predictions near relay thresholds.
\vspace{-0.8em}
\bibliographystyle{IEEEtran}
\bibliography{references}

\end{document}